\documentclass[12pt,draftclsnofoot,onecolumn, journal]{IEEEtran}

\usepackage{array}
\usepackage[ruled,vlined,linesnumbered]{algorithm2e}
\usepackage{amssymb,amsfonts,amsthm,mathrsfs}
\usepackage[cmex10]{amsmath}
\usepackage{booktabs}
\usepackage{cite}
\usepackage[usenames,dvipsnames]{color}
\usepackage{epsfig}
\usepackage{float}
\usepackage[T1]{fontenc}
\usepackage{flushend}
\usepackage{footnote}
\usepackage{graphicx}
\usepackage[latin1]{inputenc}
\usepackage{paralist}
\usepackage{subfigure}
\usepackage{setspace}
\usepackage{verbatim}
\usepackage{url,changebar,bm,xspace,dsfont}
\usepackage[normalem]{ulem}
\usepackage{xspace}
\usepackage{soul}

\makesavenoteenv{tabular}

\newcommand{\PR}{\mathds{P}}

\newcommand{\image}[3]{
\begin{figure}[!ht]
  \centering
  \includegraphics[#1]{#2}
  \caption{#3}
  \label{fig:#2}
\end{figure}
}

\newtheorem{prop}{Proposition}
\newtheorem{lem}{Lemma}

\newtheorem{remark}{Remark}

\newenvironment{list4}{
  \begin{list}{$\bullet$}{
      \setlength{\itemsep}{0.05cm}
      \setlength{\labelsep}{0.2cm}
      \setlength{\labelwidth}{0.3cm}
      \setlength{\parsep}{0in}
      \setlength{\parskip}{0in}
      \setlength{\topsep}{0in}
      \setlength{\partopsep}{0in}
      \setlength{\leftmargin}{0.17in}}}
      {\end{list}}

\begin{document}
\title{
Optimal Radio Frequency Energy Harvesting \\
     with Limited Energy Arrival Knowledge
}
\author{Zhenhua Zou, Anders Gidmark, Themistoklis Charalambous and Mikael Johansson

\thanks{
Z. Zou and T. Charalambous are with the Department of Signals and Systems, Chalmers University of Technology, Gothenburg, Sweden
(Emails:  {\tt \{zhenhua.zou,thecha\}@chalmers.se}).
A. Gidmark and M. Johansson are with the Automatic Control Lab, School of Electrical Engineering, Royal Institute of Technology (KTH), Stockholm, Sweden
(Emails:  {\tt \{gidmark,mikaelj\}@kth.se}).
}
}

\maketitle

%
%
%
%
\begin{abstract}
In this paper, we develop optimal policies for deciding when a wireless node with radio frequency (RF) energy harvesting (EH) capabilities should try and harvest ambient RF energy. While the idea of RF-EH is appealing, it is not always beneficial to attempt to harvest energy; in environments where the ambient energy is low, nodes could consume more energy being awake with their harvesting circuits turned on than what they can extract from the ambient radio signals; it is then better to enter a sleep mode until the ambient RF energy increases. Towards this end, we consider a scenario with intermittent energy arrivals and a wireless node that wakes up for a period of time (herein called the time-slot) and harvests energy. If enough energy is harvested during the time-slot, then the harvesting is successful and excess energy is stored; however, if there does not exist enough energy the harvesting is unsuccessful and energy is lost.

We assume that the ambient energy level is constant during the time-slot, and changes at slot boundaries. The energy level dynamics are described by a two-state Gilbert-Elliott Markov chain model, where the state of the Markov chain can only be observed during the harvesting action, and not when in sleep mode.
Two scenarios are studied under this model. In the first scenario, we assume that we have knowledge of the transition probabilities of the Markov chain and formulate the problem as a Partially Observable Markov Decision Process (POMDP), where we find a threshold-based optimal policy. In the second scenario, we assume that we don't have any knowledge about these parameters and formulate the problem as a Bayesian adaptive POMDP; to reduce the complexity of the computations we also propose a heuristic posterior sampling algorithm. The performance of our approaches is demonstrated via numerical examples.
\end{abstract}


\IEEEpeerreviewmaketitle

%
%
%
%
\section{Introduction}

\begin{figure}[t]
\centering
\includegraphics[width=.6\columnwidth]{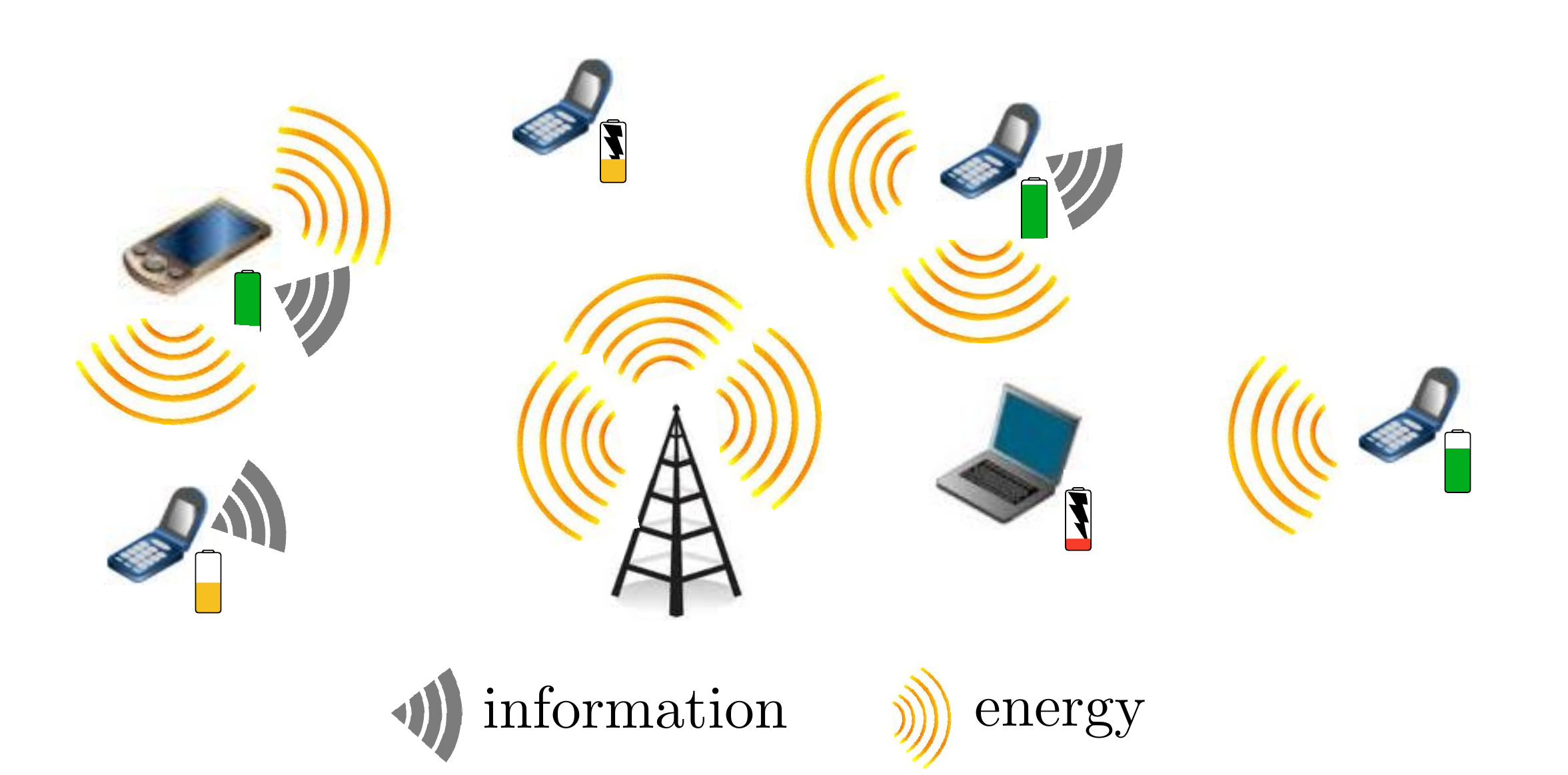}
\caption{In radio frequency energy harvesting, the device that is not the destination of the packet can capture RF radiation of the wireless transmission from cellular communication, WiFi or TV towers, and convert it into a direct current through rectennas\vspace{-0.45cm}.}
\label{fig:RF-EH}
\end{figure}

In green communications and networking, renewable energy sources can replenish the energy of network nodes and be used as an alternative power source without additional cost.
Radio frequency (RF) energy harvesting (EH) is one of the energy harvesting methods that have recently attracted a lot of attention (see, for example,~\cite{Xiao2015Survey,Ulukus2015,Ahmed2015} and references therein).
In RF-EH, a device can capture ambient RF radiation from a variety of radio transmitters (such as  television/radio broadcast stations, WiFi, cellular base stations and mobile phones), and convert it into a direct current through rectennas~\cite{rectenna}, see Figure~\ref{fig:RF-EH}.
It has been shown that low-power wireless systems such as wireless sensor networks with RF energy harvesting capabilities can have a significantly prolonged lifetime, even to the point where they can become self-sustained and support previously infeasible ubiquitous communication applications~\cite{liu2013ambient}.

However, in many cases the RF energy is intermittent. This can be due to temporary inactive periods of communication systems with bursty traffic or/and multi-path fading in wireless channels~\cite{wuperformance}. Moreover, the energy spent by wireless devices to wake up the radio and assess the channel is non-negligible. Hence, when the ambient energy is low, it is energy-inefficient for a node to try and harvest energy and better to sleep.
The challenge in the energy harvesting process lies in the fact that the wireless device does not know the energy level before trying to harvest.
For this reason, it is crucial to develop policies when a wireless node should harvest or sleep to maximize the accumulated energy.

In this paper, we study the problem of energy harvesting for a single wireless device in an environment where the ambient RF energy is intermittent. Energy harvesting with intermittent energy arrivals has been recently investigated under the scenario that the energy arrivals are described by \textit{known} Markov processes~\cite{gunduz2014designing, sharma2010optimal, michelusi2013transmission, lei2009generic, ozel2011transmission}. However, the energy arrivals may not follow the chosen Markov process model. It is therefore necessary not to presume the arrival model, but allow for an unknown energy arrival model. Towards this direction, this problem has only been targeted via the classical Q-learning method in~\cite{blasco2013learning}.
The Robbins-Monro algorithm, the mathematical cornerstone of Q-learning, was applied in~\cite{fernandezmdp2015} to derive optimal policies with a faster convergence speed by exploiting the optimal policy is threshold-based.
However, both the Q-learning method and the Robbins-Monro algorithm rely on heuristics (e.g., $\epsilon$-greedy) to handle the exploration-exploitation trade-off~\cite{watkins1992q}.
The optimal choice of the step-size for the best convergence speed is also not clear; only a set of sufficient conditions for asymptotic convergence is given.

All the aforementioned works assume that the energy arrival state is known at the decision maker, before the decision is taken. This is an unrealistic assumption since it does not take into account the energy cost for the node to wake up and track the energy arrival state, while being active continuously can be detrimental in cases of low ambient energy levels.
The partial observability issues in energy harvesting problems have only been considered in scenarios such as the knowledge of the State-of-Charge~\cite{michelusi2014optimal}, the event occurrence in the optimal sensing problem~\cite{jaggi2009rechargeable}, and the channel state information for packet transmissions~\cite{aprem2013transmit}.
To the best of our knowledge, neither the scenario with partial observability of the energy arrival nor this scenario coupled with an unknown model have been addressed in the literature before.

Due to the limited energy arrival knowledge and the cost for unsuccessful harvesting, the fundamental question being raised is whether and when it is beneficial for a wireless device to try and harvest energy from ambient energy sources.
In this paper, we aim at answering this question by developing optimal sleeping and harvesting policies that maximize the accumulated energy.
More specifically, the contributions of this paper are summarized as follows.
\begin{list4}
  \item We model the energy arrivals using an abstract two-state Markov chain model where the node receives a reward at the good state and incurs a cost at the bad state. The state of the model is revealed to the node only if it chooses to harvest.
      In absence of new observations, future energy states are predicted based on knowledge about the transition probabilities of the Markov chain.
  \item We propose a simple yet practical reward function that encompasses the effects of the decisions made based on the states of the Markov chain.
  \item We study the optimal energy harvesting problem under two assumptions on the parameters of the energy arrival model.
  \begin{enumerate}
  \item For the scenario where the parameters are known, we formulate the problem of whether to harvest or to sleep   as a Partially Observable Markov Decision Process (POMDP).
      We show that the optimal policy has a threshold structure: after an unsuccessful harvesting, the optimal action is to sleep for a constant number of time slots that depends on the parameters of the Markov chain; otherwise, it is always optimal to harvest. The threshold structure leads to an efficient computation of the optimal policy.
      Only a handful of papers have explicitly characterized the optimality of threshold-based policies for POMDP (for example,~\cite{johnston2006opportunistic,chen2009distributed}) and they do not deal with the problem considered in this work.
  \item For the scenario when the transition probabilities of the Markov chain are not known, we apply a novel Bayesian online-learning method.
      To reduce the complexity of the computations, we propose a heuristic posterior sampling algorithm.
      The main idea of Bayesian online learning is to specify a prior distribution over the unknown model parameters, and update a posterior distribution by Bayesian inference over these parameters to incorporate new information about the model as we choose actions and observe results.
      The exploration-exploitation dilemma is handled directly as an explicit decision problem modeled by an extended POMDP, where we aim to maximize future expected utility with respect to the current uncertainty on the model.
      The other advantage is that we can define an informative prior to incorporate previous beliefs about the parameters, which can be obtained from, for example, domain knowledge and field tests.
      Our work is the first in the literature that introduces and applies the Bayesian adaptive POMDP framework~\cite{ross2011bayesian} in energy harvesting problems with unknown state transition probabilities.
 \end{enumerate}
\item The schemes proposed in this paper are evaluated in simulations and significant improvements are demonstrated compared to having the wireless nodes to harvest all the time or try to harvest randomly.
\end{list4}

The rest of this paper is organized as follows. The system model and the energy harvesting problem are introduced in Section~\ref{sec:model}.
In Section~\ref{sec:pomdp} we address the case of \emph{known} Markov chain parameters, and using POMDP we derive optimal sleeping and harvesting policies; the threshold-based structure of the optimal policies are also shown.
In Section~\ref{sec:bayesian} we address the case of \emph{unknown} Markov chain parameters and we propose the Bayesian on-line learning method.
Numerical examples are provided in Section~\ref{sec:numExample}. Finally, in Section~\ref{sec:conclusions} we draw conclusions and outline possible future research directions.

%
%
%
%
\section{System Model}\label{sec:model}

We consider a single wireless device with the capability of harvesting energy from ambient energy sources. We assume that the overall energy level is constant during one time-slot, and may change in the next time-slot according to a two-state Gilbert-Elliott Markov chain model~\cite{gilbert1960capacity,elliott1963estimates}; see Fig.~\ref{fig:GE.pdf}.
\image{width=0.3\columnwidth}{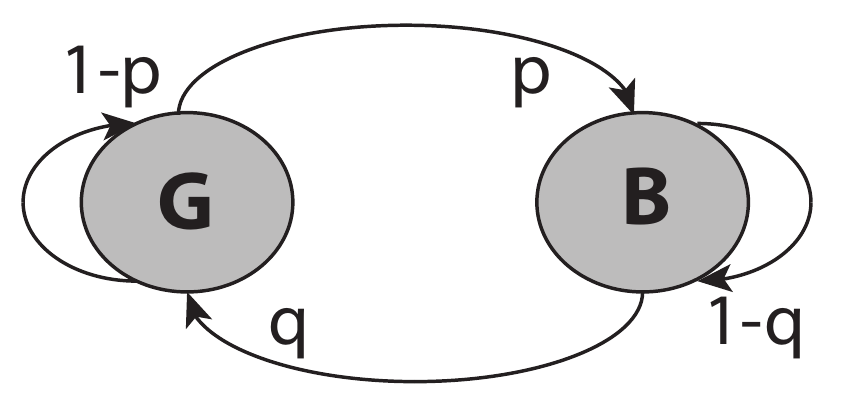}{Two-state Gilbert-Elliott Markov chain model.}
In this model, the good state~($G$) denotes the presence of energy to be harvested and the bad state~($B$) denotes the absence of energy to be harvested. The transition probability from the $G$ state to $B$ state is $p$, and the transition probability from $B$ state to $G$ state is $q$. The probabilities of staying at states $G$ and $B$ are $1-p$ and $1-q$, respectively.
It can be easily shown that the steady state distribution of the Markov chain at $B$ and $G$ states are $p/(p+q)$ and $q/(p+q)$, respectively.

At each time-slot, the node has two action choices: harvesting or sleeping. If the node chooses to harvest and the Markov chain is in the $G$ state, a reward $r_1 > 0$ is received that represents the energy successfully harvested. If the Markov chain is in the $B$ state during the harvesting action, a penalty $-r_0 < 0$ is incurred that represents the energy cost required to wake up the radio and try to detect if there exists any ambient energy to harvest.
On the other hand, if the node sleeps, no reward is received.
Therefore, the reward function is defined as
\begin{align}
R(s, a) \triangleq
\begin{cases}
r_1, & a= \mathcal{H} \wedge s = G, \\
-r_0, & a= \mathcal{H} \wedge s = B,  \\
0, & a = \mathcal{S},
\end{cases}
\label{eq:Reward}
\end{align}
where $a$ denotes the harvesting action ($\mathcal{H}$) or the sleeping action ($\mathcal{S}$), and $s$ is the current state of the Markov chain.

\begin{remark}
Note that one could impose a cost for sleeping. However, this does not change the problem setup since we could normalize the rewards and costs so that the sleeping cost is zero.
\end{remark}

\begin{remark}
In addition, the choice of the exact numbers for $r_0$ and $r_1$ depend on hardware specifications, such as the energy harvesting efficiency and the energy harvesting cost. Even though in reality the energy harvested and hence the reward $r_1$ is not fixed, the choice of $r_1$ can be seen as the minimum or average energy harvested during a time-slot. Similarly, $r_0$ can be seen as the maximum or average energy spent during a slot when the node failed to harvest energy.
\end{remark}

The state information of the underlying Markov chain can only be observed by the harvesting action, but there is a cost associated with an unsuccessful energy harvesting. On the other hand, sleeping action neither reveals the state information nor incurs any cost. Thus, it is not immediately clear when it is better to harvest to maximize the reward. Furthermore, the transition probabilities of the Markov chain may not be known a priori, which makes the problem of maximizing the reward even more challenging.

Let $a_t \in \{\mathcal{H}, \mathcal{S}\}$ denote the action at time $t$, $s_t$ denote the state of the Markov chain at time $t$, and $z_t \in \{G, B, Z\}$ denote the observation at time $t$ where $Z$ means no observation of the Markov chain.
Let $a^t \triangleq \{a_0, a_1, \dotsc, a_t\}$ denote the history of actions and $z^t \triangleq \{z_0, z_1, \dotsc, z_t\}$ denote the history of observations.
A policy $\pi$ is a function that randomly prescribes an action at time $t$ based on the history of actions and observations up to time $t-1$.
The goal is then to find the optimal policy $\pi^\star$ that maximizes the expected total discounted reward,
\[
\pi^\star \in \arg \max_{\pi} \mathds{E}_\pi [\sum_{t=0}^{\infty} \gamma^t R_t (s_t,a_t)] ,
\]
where $R_t$ is the reward at time $t$ and the expectation is taken with respect to the randomization in the policy and the transitions of the Markov chain.
The discount factor $\gamma \in [0,1)$ models the importance of the energy arrivals at different time slots in which the energy harvested in the future will be discounted.
The discount factor can also be seen as a scenario where the node terminates its operation at each time-slot independently with probability $(1-\gamma)$~\cite{Puterman2005}.

%
%
%
%
\section{Optimal structured policy with unknown Markovian states}\label{sec:pomdp}

In this section, we first solve the problem of deriving the optimal policy with \textit{known} transition probabilities and \textit{unknown} Markovian states by formulating it as a Partially Observable Markov Decision Process (POMDP)~\cite{kaelbling1998planning}.
We further show that the optimal policy has a threshold-based structure. This structural result simplifies both the off-line computations during the design phase and the real-time implementation.

%
%
\subsection{POMDP formulation}

Although the exact state is not known at each time-slot, we can keep a probability distribution (i.e.,~belief) of the state based on the past observations and the knowledge on the Markov chain. It has been shown that such a belief is a sufficient statistic~\cite{kaelbling1998planning}, and we can convert the POMDP to a corresponding MDP with the belief as the state.

Let the scalar $b$ denote the belief that the state is good (i.e., $G$) at the current time-slot. If the action is to harvest at the current time-slot, in the next time-slot the belief can be either $b_B \triangleq q$ or $b_G \triangleq 1-p$ depending on the harvesting result. If the action is to sleep, the belief is updated according to the Markov chain, i.e.,
\begin{align}\label{eq:b_slash}
b'= (1-p)b + q(1-b) = q + (1-p-q)b,
\end{align}
which is the probability of being at good state at the next time-slot given the probability at the current time-slot.
This update converges to the stationary distribution of the good state. In summary, we have the following state transition probability
\begin{align*}
\PR(b'|a,b) =
\begin{cases}
b & \text{if } a = \mathcal{H}, b' = b_G, \\
1-b & \text{if } a = \mathcal{H}, b' = b_B, \\
1 & \text{if } a = \mathcal{S}, b' = q + (1-p-q)b, \\
0 & \text{otherwise.}
\end{cases}
\end{align*}
We let $1-p > q$, which has the physical meaning that the probability of being at $G$ state is higher if the state at the previous time is in $G$ state other than in $B$ state. Please let me know if the grammar of this sentence is correct or not.
Hence, the belief $b$ takes discrete values between $q$ and $1-p$, and the number of belief is infinite but countable.

By Equation~\eqref{eq:Reward}, the expected reward with belief $b$ is
\begin{align*}
R(b, a)
&= b R(1,a) + (1-b) R(0,a) \\
& =
\begin{cases}
(r_0 + r_1) b - r_0, & a = \mathcal{H}, \\
0, & a = \mathcal{S}.
\end{cases}
\end{align*}

Any combination of the action history $a^t$ and the observation history $z^t$ corresponds to a unique belief $b$.
Hence, the policy $\pi$ is also a function that prescribes a random action $a$ for the belief $b$.
The expected total discounted reward for a policy $\pi$ starting from initial belief $b_0$, also termed as the value function, is then
\[
V^\pi(b_0) \triangleq \mathds{E}_\pi [\sum_{t=0}^{\infty} \gamma^t R_t(b_t,a_t)].
\]
Since the state space is countable and the action space is finite with only two actions, there exists an optimal deterministic stationary policy
$\pi^\star$ for any $b$~\cite[Theorem 6.2.10]{Puterman2005} such that
\[
\pi^\star \in \arg \max_{\pi} V^\pi(b).
\]

%
%
\subsection{Optimal policy - value iteration}

Let $V^\star \triangleq V^{\pi^\star}$ be the optimal value function. The optimal policy can be derived from the optimal value function, i.e.,
for any $b$, we have
\begin{align*}
& \pi^\star(b) \in \arg
  \max_{a \in \{\mathcal{H}, \mathcal{S} \}}
\big[ R(b,a) + \gamma \sum_{b'} \PR(b'|a,b) V^\star(b') \big].
\end{align*}

The problem of deriving the optimal policy is then to compute the optimal value function.
It is known that the optimal value function satisfies the Bellman optimality equation~\cite[Theorem 6.2.5]{Puterman2005},
\[
V^\star(b) =
\max_{a \in \{\mathcal{H}, \mathcal{S} \}}
\big[ R(b,a) + \gamma \sum_{b'} \PR(b'|a,b) V^\star(b') \big],
\]
and the optimal value function can be found by the value iteration method shown in Algorithm~\ref{alg:ValueIteration}.
The algorithm utilizes the fixed-point iteration method to solve the Bellman optimality equation with stopping criteria.
If we let $t \rightarrow \infty$, then the algorithm returns the optimal value function $V^\star(b)$~\cite{Puterman2005}.
\begin{algorithm}
\SetAlgoLined
\DontPrintSemicolon
    \KwIn{Error bound $\epsilon$}
    \KwOut{$V(b)$ with $\sup_{b}|V(b) - V^\star(b)| \leq \epsilon / 2$.}
    Initialization: At $t = 0$, let $V_0(b) = 0$ for all $b$\;
    \Repeat{$\sup_{b}|V_{t+1}(b) - V_t(b)| \leq \epsilon (1 - \gamma) / 2 \gamma.$}{
    Compute $V_{t+1}(b)$ for all states $b$,
    \[
      V_{t+1}(b) = \max_{a \in \{\mathcal{H}, \mathcal{S} \} }
      \big[ R(b,a) + \gamma \sum_{b'} \PR(b'|a,b) V_t(b') \big].
    \]
    Update $t = t + 1.$
    }
\caption{Value iteration algorithm~\cite{Puterman2005}}
\label{alg:ValueIteration}
\end{algorithm}

%
%
\subsection{Optimality of the threshold-based policy}
Let $V_{t+1}(b,a)$ denote the value function of any action $a \in \{\mathcal{H}, \mathcal{S}\}$ in Algorithm~\ref{alg:ValueIteration}, and let $V_{\infty}(b,a) = \lim_{t \rightarrow \infty} V_t (b,a).$ We first show that the optimal policy has a threshold structure:
\begin{prop}
\label{prop:waitTime}
Define
\[
\overline{b} \triangleq
\min_{b} \{ V_{\infty}(b,\mathcal{H}) \geq V_{\infty}(b,\mathcal{S}) \}.
\]
If the threshold $\overline{b} \geq q/(p+q)$, then the optimal policy is to never harvest.
If  $\overline{b} < q/(p+q)$, then the optimal policy is to continue to harvest after a successful harvesting time slot, and to sleep for
\[
N \triangleq \left\lceil \log_{1-p-q}\dfrac{q - (p+q)\overline{b}}{q} \right\rceil - 1
\]
time slots after an unsuccessful harvesting.
\end{prop}
\begin{IEEEproof}
The proof relies on two Lemmas presented in the end of this section.
We first prove that the optimal action is to harvest for any belief $b \geq \overline{b}$ and to sleep for any belief $b < \overline{b}$.
From the definition of $\overline{b}$, it is clear that it is always optimal to sleep for belief $b < \overline{b}$.
From Equation~\eqref{eq:sleepUpdate} and Equation~\eqref{eq:senseUpdate}, we have that
\begin{align*}
V_{\infty}(b,\mathcal{H}) &= \alpha_{h,\infty} + \beta_{h,\infty} b, \\
V_{\infty}(b,\mathcal{S}) &= \max_{ \{\alpha_s, \beta_s\} \in \Gamma_{s,\infty}}
\{\alpha_s + \beta_s b \}
\end{align*}
where $\Gamma_{s, \infty} =
\{ \gamma (\alpha+\beta q), \gamma \beta(1-p-q)
: \forall \{\alpha, \beta\}
\in \Gamma_{\infty} \}$,
and $\Gamma_{\infty} = \Gamma_{s,\infty} \bigcup \{\alpha_{h,\infty}, \beta_{h,\infty}\}$.
Let $B_{s,\infty} \triangleq \{\beta: \{\alpha,\beta\} \in \Gamma_{s,\infty}\}$
and $B_{\infty} \triangleq B_{s,\infty} \bigcup \beta_{h,\infty}.$
Hence, every $\beta$ value in $B_{s,\infty}$ is generated by a scaling factor $\gamma(1-p-q)$ from the set $B_{\infty}$.
Since $\gamma(1-p-q)$ is strictly smaller than one and $\beta \geq 0$ from Lemma~\ref{lem:monotoneOptimalValue}, we have that
$\beta_{h,\infty} \geq \max\{\beta_s\}$ by the proof of contradiction.
Since $V_{\infty}(\overline{b},\mathcal{H}) \geq V_{\infty}(\overline{b},\mathcal{S})$, it follows that $V_{\infty}(b,\mathcal{H}) \geq V_{\infty}(b,\mathcal{S})$ for any $b \geq \overline{b}$.

Observe that after an unsuccessful harvesting and sleeping additionally for $t-1$ time slots, the belief $b$ is
\[
q \sum_{i=0}^{t-1}(1-p-q)^i = q \dfrac{1-(1-p-q)^t}{p+q}.
\]
Since $1-p-q \in (0, 1)$, this is monotonically increasing with $t$ and converges to $q/(p+q)$.
The proposition follows by deriving $t$ such that the belief is larger than the threshold $\overline{b}$.
\end{IEEEproof}

Proposition~\ref{prop:waitTime} suggests that we can focus on the set of policies with threshold-structure, which is a much smaller set than the set of all policies.
This leads to an efficient computation of the optimal policy shown in Proposition~\ref{prop:efficientComp}.
\begin{prop}
\label{prop:efficientComp}
Let $b' \triangleq q[1-(1-p-q)^{n+1}]/(p+q)$,
let $F(n) \triangleq \gamma^{n+1} r_1 (b' - 1 + p) + r_1 - p(r_0 + r_1)$,
and let $G(n) \triangleq \gamma^{n+1}(b'(1-\gamma) - (1-\gamma+\gamma p)) + 1 - \gamma + \gamma p$.
The optimal policy is to continuously harvest after a successful harvesting, and to sleep for
\[
N \triangleq {\arg \max}_{n \in \{0, 1, \dotsc\}} \dfrac{F(n)}{G(n)}
\] time slots after an unsuccessful harvesting.
\end{prop}
\begin{IEEEproof}
Let $\pi^n$ denote the policy that sleeps $n$ time slots after bad state observation, and always harvests after good state observation. By Proposition~\ref{prop:waitTime}, the optimal policy is a type of $\pi^n$ policy, and we need to find the optimal sleeping time that gives the maximum reward.

Recall that the belief after good state observation is $1-p$, and after bad state observation is $q$.
The belief after bad state observation and sleeping $n$ time slots is
\[
b' \triangleq q \sum_{i=0}^{n}(1-p-q)^i = q \dfrac{1-(1-p-q)^{n+1}}{p+q}.
\]
At belief $q$, the $\pi^n$ policy is to sleep for $n$ time slots, and thus
\begin{align}
V^{\pi^n}(q) = \gamma^n V^{\pi^n}(b').
\label{eq:derivePolicy1}
\end{align}
At belief $1-p$, the $\pi^n$ policy is to harvest, and thus
\begin{align}
\nonumber V^{\pi^n}(1-p) &= (1-p)(r_0+r_1) - r_0 \\
& \qquad + \gamma p V^{\pi^n}(q) + \gamma(1-p)V^{\pi^n}(1-p).
\label{eq:derivePolicy2}
\end{align}
At belief $b'$, the $\pi^n$ policy is also to harvest, and thus
\begin{align}
\nonumber V^{\pi^n}(b') &= b'(r_0+r_1) - r_0 \\
 & \qquad + \gamma^{n+1} (1-b') V^{\pi^n}(b') +
 \gamma b' V^{\pi^n}(1-p).
\label{eq:derivePolicy3}
\end{align}
By solving the above Equations~\eqref{eq:derivePolicy1}-\eqref{eq:derivePolicy2}-\eqref{eq:derivePolicy3}, $V^{\pi^n}(1-p)$ corresponds to $F(n)/G(n)$.
Hence, $N$ is the optimal sleeping time that gives the maximum reward within the set of policies defined by $\pi^n$.
Since the optimal policy has this structure, the proposition is then proved.
\end{IEEEproof}

\begin{lem}
The value function $V_t(b)$ in the value iteration algorithm at any time $t$ is a piecewise linear convex function over belief $b$, i.e.,
\[
V_t(b) =
\max_{\{\alpha,\beta\} \in \Gamma_t \subset \mathds{R}^2 } \{ \alpha + \beta b \} ,
\]
where the set $\Gamma_t$ is computed iteratively from the set $\Gamma_{t-1}$ with the initial condition $\Gamma_0 = \{0, 0\}$.
\end{lem}
\begin{IEEEproof}
We prove the lemma by induction on time $t$.
The statement is correct when $t=0$ with $\Gamma_0 = \{0,0\}$ since $V_0(b) = 0$ for all $b$.
Suppose the statement is correct for any $t$. The value function of sleeping action at time $t+1$ is
\begin{align*}
V_{t+1}(b, \mathcal{S})
& \triangleq \gamma V_t(q + b (1-p-q)) \\
&= \gamma  \max_{\{\alpha, \beta\} \in \Gamma_t}
 \{\alpha + \beta  (q + b  (1-p-q)) \} \\
& = \max_{\{\alpha, \beta\} \in \Gamma_t}
 \{ \gamma  ( \alpha+\beta q) + b \gamma \beta  (1-p-q) \} .
\label{eq:sleepUpdate}
\end{align*}
Define
\begin{align*}
\Gamma_{s,t+1} &\triangleq \{ \gamma  (\alpha+\beta q), \gamma \beta  (1-p-q): \forall \{\alpha, \beta\} \in \Gamma_t \},\\
\alpha_s &\triangleq \gamma  (\alpha+\beta q),\\
\beta_s &\triangleq \gamma \beta  (1-p-q).
\end{align*}
Hence, we have
\begin{align}
V_{t+1}(b, \mathcal{S}) = \max_{ \{\alpha_s, \beta_s\} \in \Gamma_{s,t+1}}
\{\alpha_s + \beta_s b \}.
\end{align}

\noindent The value function of the harvesting action is
\begin{align*}
V_{t+1}(b,\mathcal{H}) & \triangleq \; (r_0 + r_1)b - r_0+ \gamma V_{t}(b_B)  (1- b) + \gamma V_{t}(b_G)  b \\
=  -&r_0 +  \gamma V_{t}(b_B)+ (r_0 + r_1 + \gamma (V_{t}(b_G) - V_{t}(b_B))) b .
\end{align*}
Define
\begin{align*}
\alpha_{h,t} &\triangleq -r_0 + \gamma V_{t}(b_B),\\
\beta_{h,t} &\triangleq r_0 + r_1 + \gamma (V_{t}(b_G) - V_{t}(b_B)).
\end{align*}
We then have
\begin{align}
\label{eq:senseUpdate}
  V_{t+1}(b,\mathcal{H}) = \; & \alpha_{h,t} + \beta_{h,t} b.
\end{align}
Since $V_{t+1}(b) = \max\{V_{t+1}(b,\mathcal{S}), V_{t+1}(b,\mathcal{H})\}$, the statement is proved by defining $\Gamma_{t+1} \triangleq \{ \alpha_{h,t}, \beta_{h,t} \} \bigcup \Gamma_{s,t+1}$.
\end{IEEEproof}

\begin{lem}
\label{lem:monotoneOptimalValue}
For any $t$, if $b_1 \geq b_2$, then $V_t(b_1) \geq V_t(b_2)$.
For any $\{\alpha,\beta\} \in \Gamma_{t}$, we have $\beta \geq 0$.
\end{lem}
\begin{IEEEproof}
We prove the proposition by induction on time $t$.
Since $V_0(b) = 0$ for all $b$ at time $t=0$ and $\Gamma_0 = \{0,0\}$, the statement is correct at time $t=0$.
Suppose the statement is correct at time $t$.
Since $1-p-q \geq 0$ and $\beta \geq 0$, we have that
\[
\gamma (\alpha+\beta q) + b_1 \gamma \beta (1-p-q) \geq
\gamma (\alpha+\beta q) + b_2 \gamma \beta (1-p-q).
\]
By Equation~\eqref{eq:sleepUpdate}, we have $V_{t+1}(b_1, \mathcal{S}) \geq V_{t+1}(b_2, \mathcal{S})$.
Since $b_G > b_B$, we also have $V_t(b_G) \geq V_t(b_B)$ by the induction condition.
By Equation~\eqref{eq:senseUpdate}, we have $V_{t+1}(b_1, \mathcal{H}) \geq V_{t+1}(b_2, \mathcal{H})$.
Hence, we have that $V_{t+1}(b_1) \geq V_{t+1}(b_2)$.
Similarly, we can also derive that $\beta \geq 0$ for any $\{\alpha, \beta\} \in \Gamma_{t+1}$.
\end{IEEEproof}

%
%
%
%
\section{Bayesian online learning unknown transition probabilities}\label{sec:bayesian}

In many practical scenarios, the transition probabilities of the Markov chain that model the energy arrivals may be initially unknown. To obtain an accurate estimation, we need to sample the channel many times, a process which unfortunately consumes a large amount of energy and takes a lot of time. Thus, it becomes crucial to design algorithms that balance the parameter estimation and the overall harvested energy; this is the so-called exploration and exploitation dilemma. Towards this end, in this section, we first formulate the optimal energy harvesting problem with unknown transition probabilities as a Bayesian adaptive POMDP~\cite{ross2011bayesian}. Next, we propose a heuristic posterior sampling algorithm based on the threshold structure of the optimal policy with known transition probabilities. The Bayesian approach can incorporate the domain knowledge by specifying a proper prior distribution of the unknown parameters. It can also strike a natural trade-off between exploration and exploitation during the learning phase.

\subsection{Models and Bayesian update}
The Beta distribution is a family of distributions that is defined on the interval $[0, 1]$ and parameterized by two parameters.
It is typically used as conjugate prior distributions for Bernoulli distributions so that the posterior update after observing state transitions is easy to compute.
Hence, for this work, we assume that the unknown transition probabilities $p$ and $q$ have independent prior distributions following the Beta distribution parameterized by $\phi \triangleq [\phi_1 \ \phi_2 \ \phi_3 \ \phi_4]^T \in \mathds{Z}_{+}^4$, i.e.,
\begin{align*}
\PR(p, q ;\phi)  & = \PR(p,q; \phi_1, \phi_2, \phi_3, \phi_4) \\
&\stackrel{(a)}{=} \PR(p; \phi_1, \phi_2) \PR(q; \phi_3, \phi_4) ,
\end{align*}
\noindent where $(a)$ stems from the fact that $p$ and $q$ have independent prior distributions. The Beta densities of probabilities $p$ and $q$ are given by
\begin{align*}
\PR(p; \phi_1, \phi_2) &= \dfrac{\Gamma(\phi_1 + \phi_2) }{ \Gamma(\phi_1) \Gamma(\phi_2)} p^{\phi_1 - 1} (1-p)^{\phi_2 - 1}, \\
\PR(q; \phi_3, \phi_4)&= \dfrac{\Gamma(\phi_3 + \phi_4)}{\Gamma(\phi_3) \Gamma(\phi_4)}{q^{\phi_3 - 1} (1-q)^{\phi_4 - 1}},
\end{align*}
respectively, where $\Gamma(\cdot)$ is the gamma function, given by $\Gamma(y)=\int^{\infty}_{0} x^{y-1}e^{-x}dx$.
However, for $y\in \mathbb{Z}_{+}$ (as it is the case in our work), the gamma function becomes $\Gamma(y)=(y-1)!$.

By using the Beta distribution parameterized by posterior counts for $p$ and $q$, the posterior update after observing state transitions is easy to compute.
For example, suppose the posterior count for the parameter $p$ is $\phi_1 = 5$ and $\phi_2 = 7$.
After observing state transitions from $G$ to $B$ (with probability $p$) for $2$ times and state transitions from $G$ to $G$ (with probability $1-p$) for $3$ times, the posterior count for the parameter $p$ is simply $\phi_1 = 5 + 2 = 7$ and $\phi_2 = 7 + 3 = 10$.
Without loss of generality, we assume that $\phi$ initially is set to $[1,1,1,1]$ to denote that the parameters $p$ and $q$ are between zero and one with equal probabilities.

Note that we can infer the action history $a^t$ from the observation history $z^t$.
More specifically, for each time $t$, if $z_t = Z$, then $a_t = \mathcal{S}$, and if $z_t \in \{G, B\}$, then $a_t = \mathcal{H}$.
In what follows, we use only the observation history $z^t$ for posterior update for the sake of simplicity.
Consider the joint posterior distribution $\PR(s_t, p, q | z^{t-1})$ of the energy state $s_t$ and the transition probability $p$ and $q$ at time $t$ from the observation history $z^{t-1}$.
Let
$$
S(z^{t-1}) = \{s^{t-1} : s_\tau = z_\tau \; \forall \tau \in \{t' : z_{t'} \neq Z\}\}
$$
denote all possible state history based on the observation history $z^{t-1}$.
Let $C(\phi, S(z^{t-1}), s_t)$ denote the total number of state histories that lead to the posterior count $\phi$ from the initial condition that all counts are equal one, and we call it the \textit{appearance count} to distinguish from the posterior count $\phi$.
Hence,
\begin{align*}
\PR&(s_t, p, q |z^{t-1}) \PR(z^{t-1}) \\
& \hspace{-5mm} = \PR (z^{t-1}, s_t|p, q) \PR(p,q) =
\sum_{s^{t-1}}
\PR (z^{t-1}, s^t|p, q) \PR(p,q)  \\
& \hspace{-5mm} =  \sum_{s^{t-1} \in S(z^{t-1})}
  \PR (s^t|p, q) \PR(p,q)  \\
& \hspace{-5mm} = \sum_{\phi} C(\phi, S(z^{t-1}), s_t)
p^{\phi_1 - 1} (1-p)^{\phi_2 - 1}
q^{\phi_3 - 1} (1-q)^{\phi_4 - 1},
\end{align*}
which can be written as
\begin{align*}
\PR(s_t, p, q |z^{t-1})
& \triangleq \sum_{\phi} \PR(\phi,s_t | z^{t-1}) \PR(p,q|\phi),
\end{align*}
where
\[
\PR(\phi,s_t | z^{t-1}) \triangleq
\dfrac{C(\phi, S(z^{t-1}), s_t)
\Pi_{i=1}^{4}\Gamma(\phi_i)}
{\PR(z^{t-1})\Gamma(\phi_1 + \phi_2) \Gamma(\phi_3 + \phi_4)}.
\]
Therefore, the posterior $\PR(s_t, p, q |z^{t-1})$ can be seen as a probability distribution over the energy state $s_t$ and the posterior count $\phi$.
Furthermore, the posterior can be fully described by each appearance count $C$ associated with the posterior count $\phi$ and the energy state $s_t$, up to the normalization term $\PR(z^{t-1})$.

When we have a new observation $z_t$ at time $t$, the posterior at time $t+1$ is updated in a recursive form as follows
\begin{align*}
\PR&(s_{t+1}, p, q | z^{t}) = \PR(s_{t+1}, p, q | z^{t-1}, z_{t}) \\
&= \sum_{s_t} \PR(s_t, p, q, s_{t+1} | z^{t-1}, z_{t}) \\
&= \sum_{s_t} \PR(s_t, p, q, s_{t+1}, z_{t}| z^{t-1}) / \PR(z_t | z^{t-1}) \\
&= \sum_{s_t} \PR(s_t, p, q| z^{t-1}) \PR(s_{t+1}, z_{t}|s_t, p, q, z^{t-1})
 / \PR(z_t | z^{t-1}) \\
&= \sum_{s_t} \PR(s_t, p, q| z^{t-1}) \PR(s_{t+1}, z_{t}|s_t, p, q)
 / \PR(z_t | z^{t-1}),
\end{align*}
where $\PR(z_t | z^{t-1})$ is the normalization term.

If we harvest and observe the exact state, the total number of possible posterior counts will remain the same.
For example, if we harvest and observe that $z_t = G$, this implies that $s_t = G$.
The posterior for $s_{t+1} = B$ is then
\begin{align*}
\PR&(B, p, q | z^{t}) \PR(z_t | z^{t-1})\\
&= \PR(G, p, q| z^{t-1}) \PR(B|G, p, q)  \\
&= \sum_{\phi} \PR(\phi,G | z^{t-1}) \PR(p,q|\phi_1+1,\phi_2,\phi_3,\phi_4).
\end{align*}
This update has the simple form that we take the posterior count $\phi$ associated with $G$ state at the previous update, and increase the posterior count $\phi_1$ by one.
On the other hand, the total number of possible posterior counts will be at most multiplied by two for the sleeping action.
For example, if the action is to sleep, i.e., $z_t = Z$, then we have to iterate over two possible states at time $t$ since we do not know the exact state.
The posterior for $s_{t+1} = B$ is then
\begin{align*}
\PR&(B, p, q | z^{t}) \PR(z_t | z^{t-1}) \\
&= \sum_{s_t \in \{G,B\}} \PR(s_t, p, q| z^{t-1}) \PR(B|s_t, p, q) \\
&= \big[ \sum_{\phi} \PR(\phi,G | z^{t-1})
\PR(p,q|\phi_1+1,\phi_2,\phi_3,\phi_4) \\
& \quad + \sum_{\phi} \PR(\phi,B | z^{t-1})
\PR(p,q|\phi_1,\phi_2,\phi_3,\phi_4+1) \big].
\end{align*}
The updates in other scenarios can be defined similarly.
An example of the update of the appearance count is shown in Figure~\ref{fig:beliefUpdate.pdf}.
\image{width=0.8\columnwidth}{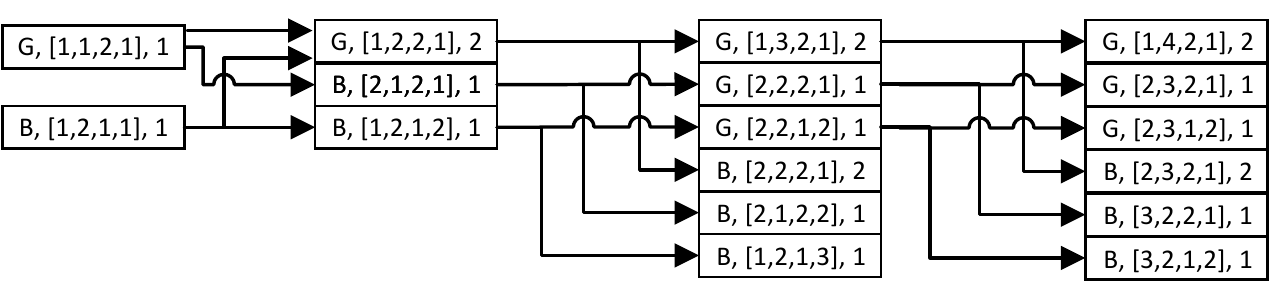}{A belief-update example after \textit{two sleeping actions} and \textit{one harvesting action with good state observation}. The numbers in the rectangle denote respectively the energy state ($G$ or $B$), the posterior count $\phi$ and the appearance count $C$.}
Note that two previously different posterior counts could lead to the same value after one update, in which we simply add their appearance count.

%
%

\subsection{Extended POMDP formulation of the Bayesian framework}
The problem is then to derive an optimal policy in order to maximize the expected reward based on the current posterior distribution of the energy states and the state transition probabilities, obtained via the Bayesian framework described. This has been shown to be equivalent to deriving an optimal policy in an extended POMDP~\cite{ross2011bayesian}.

In what follows, we will show the detailed formulation of the POMDP.
In the POMDP, the state space is $\{G,B\} \times \mathds{Z}_{+}^4$ that denotes the energy state and the posterior count $\phi$ of the Beta distribution.
The action space and the reward function do not change.
For brevity, we let $I_{t} \triangleq \{s_{t-1}, \phi, a_{t} \}$.
Recall that the state of this POMDP is $\{s_{t-1}, \phi\}$.
By the formula of conditional probability and the independence assumptions, the joint state transition and observation probability~is
\begin{align*}
\PR(s_{t}, \phi', z_{t} | I_t )& = \PR(s_{t} |I_t) \PR(z_{t} |I_t, s_{t}) \PR(\phi'|I_t, s_{t}, z_{t}) \\
&= \PR(s_{t} |s_{t-1}, \phi)
\PR(z_{t} | s_{t}) \PR(\phi'|s_{t-1}, \phi, s_{t}),
\end{align*}
where $\PR(z_{t} | s_{t}) = 1$ if $z_t = s_t$, and
$\PR(\phi' | s_{t-1}, \phi, s_t) = 1$ if the change of state from $s_{t-1}$ to $s_{t}$ leads to the corresponding update of $\phi$ to $\phi'$.
Lastly, the transition $\PR(s_{t} |s_{t-1}, \phi)$ is derived from the average $p$ and $q$ associated with the posterior count $\phi$. For example, if $s_{t-1} = G$ and $s_{t} = B$, then
$
\PR(s_{t} |s_{t-1}, \phi) = \phi_1 / (\phi_1 + \phi_2).
$
Therefore, the problem of deriving the optimal policy in the Bayesian framework can be solved by techniques developed for the POMDP.
The optimal policy tackles the exploration and exploitation dilemma by incorporating the uncertainty in the transition probabilities in the decision making processes.

%
%
\subsection{Heuristic learning algorithm based on posterior sampling}

It is computationally difficult to solve the extended POMDP exactly due to its large state space. More precisely, during the Bayesian update, we keep the appearance count of all the possible posterior count $\phi$ and the energy state ($G$ or $B$). The challenge is that the number of possible posterior count $\phi$ is multiplied by two after the sleeping action, and it can grow to infinity.
One approach could be to ignore the posterior update with the sleeping action, and the number of posterior count is kept constant at two.
However, this approach is equivalent to heuristically assuming that the unknown energy state is kept the same during the sleeping period.

Instead, we propose the heuristic posterior sampling algorithm~\ref{alg:Heuristic} inspired by~\cite{ross2011bayesian, strens2000bayesian}.
The idea is to keep the $K$ posterior counts that have the largest appearance count in the Bayesian update.
If the energy state was in good state, then we keep harvesting.
If the energy state was in bad state, then we get a sample of transition probabilities from the posterior distributions, and find the optimal sleeping time corresponding to the sampled transition probabilities.
The idea leverages on the fact the optimal policy with respect to a given set of transition probabilities is threshold-based and can be pre-computed off-line.

More precisely, the algorithm maintains the value $\psi^G \triangleq [\phi_1,\phi_2,\phi_3,\phi_4,n]$ that denotes the appearance count $n$ that leads to the posterior count $[\phi_1,\phi_2,\phi_3,\phi_4]$ and the good state.
The value $\psi^B$ is defined similarly.
The two procedures in Line~\ref{algLine:goodStateUpdate} and Line~\ref{algLine:badtateUpdate} show the computation of the update of the posterior count and appearance count with good and bad state observations, respectively.
We uniformly pick a posterior count according to their appearance counts shown in Line~\ref{algLine:PosteriorSample1} to reduce computational complexity.
The transition probability is taken to be the mean of the Beta distribution corresponding to the sampled posterior count as shown in Line~\ref{algLine:PosteriorSample2}.
Lastly, with the sleeping action, we have to invoke both good state and bad state updates in Line~\ref{algLine:sleepUpdate1}~and~\ref{algLine:sleepUpdate2}, since the state is not observed.

\begin{algorithm}
\SetAlgoLined
\DontPrintSemicolon
\SetKwFunction{GoodState}{Good State Update}
\SetKwFunction{BadState}{Bad State Update}
    \KwIn{$r$, $\gamma, K$, optimal policy lookup table}
    Initialization: Let sleeping time $w = 0$ \\
    \While{true}{
    \eIf{sleeping time $w = 0$}
    {
    Harvest energy \;
    \eIf{Successfully with good state}
    {
    \GoodState{} \;
    Sleeping time $w = 0$
    }{
    \BadState{} \;
    \label{algLine:PosteriorSample1} Draw $\psi^G$ or $\psi^B$ proportional to the count $n$ \;
    \label{algLine:PosteriorSample2} Let $p = \phi_1/(\phi_1 + \phi_2), q = \phi_3/(\phi_3 + \phi_4)$ \;
    Find sleeping time $w$ from the lookup table
    }
    }{
    Sleep and decrease sleeping time $w = w - 1$ \;
    \label{algLine:sleepUpdate1} \GoodState{} \;
    \label{algLine:sleepUpdate2} \BadState{} \;
    }
    Merge $\overline{\psi^G}$ and $\overline{\psi^B}$ with same posterior count by summing appearance count $n$ \;
    Assign $2K$ items of $\overline{\psi^G}$ and $\overline{\psi^B}$ with the highest number of $n$ to $\psi^G$ and $\psi^B$, respectively.
    }
  \SetKwProg{myprocG}{Procedure}{}{}
  \myprocG{\GoodState{} \label{algLine:goodStateUpdate} }{
  For each $\psi^G$, generate new $\overline{\psi^G}$ such that $\overline{\psi^G}(\phi_2) = \psi^G(\phi_2) + 1$
  and new $\overline{\psi^B}$ such that
  $\overline{\psi^B}(\phi_1) = \psi^G(\phi_1) + 1$ \;
  }
  \SetKwProg{myprocB}{Procedure}{}{}
  \myprocB{\BadState{} \label{algLine:badtateUpdate} }{
  For each $\psi^B$, generate new $\overline{\psi^G}$ such that $\overline{\psi^G}(\phi_3) = \psi^G(\phi_3) + 1$
  and new $\overline{\psi^B}$ such that
  $\overline{\psi^B}(\phi_4) = \psi^G(\phi_4) + 1$ \;
  }
\caption{Posterior-sampling algorithm}
\label{alg:Heuristic}
\end{algorithm}

%
%
%
%
\section{Numerical Examples}
\label{sec:numExample}

\subsection{Known transition probabilities}

\image{width=0.5\columnwidth}{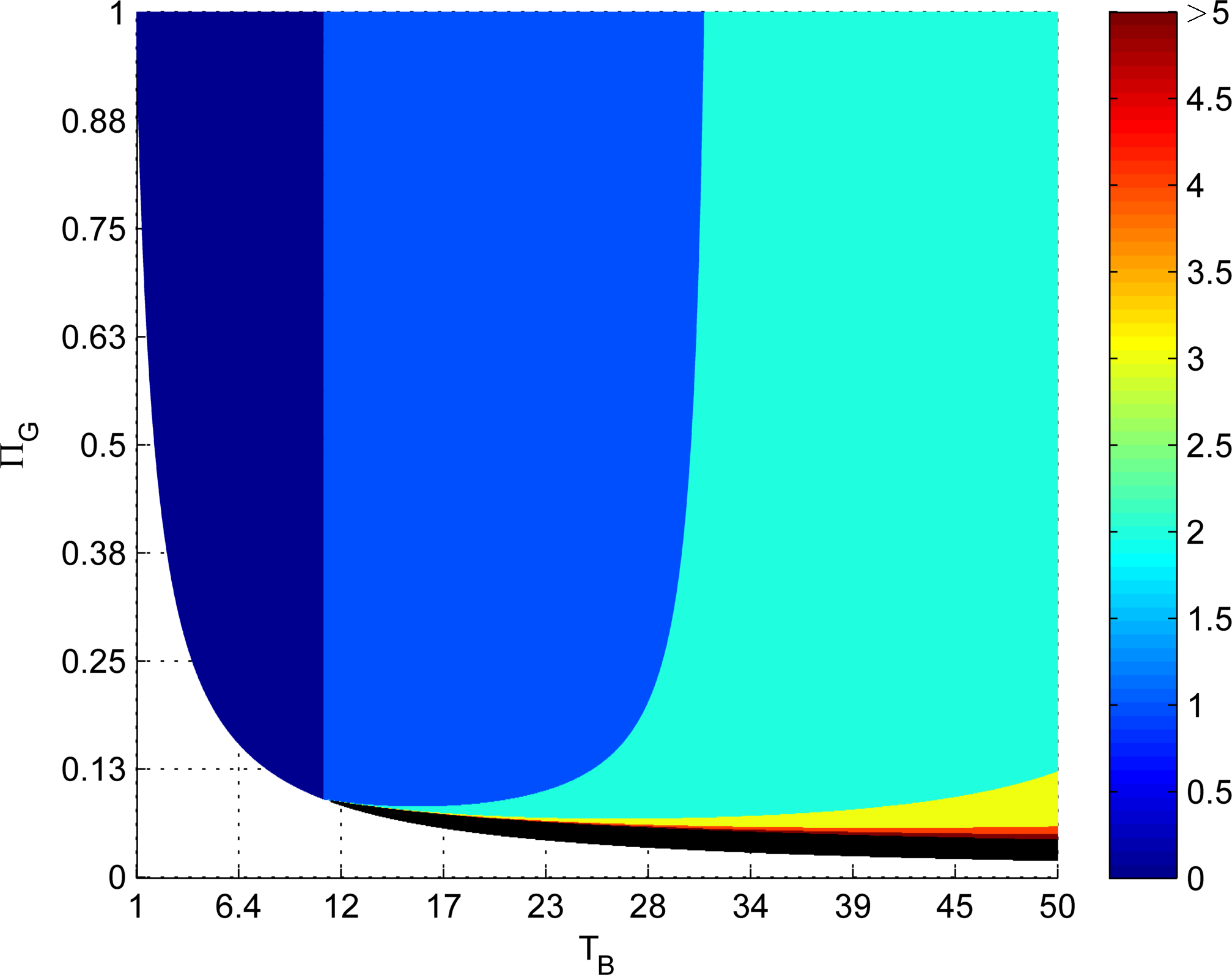}{Optimal sleeping time with $r_1 = 10$, $r_0 = 1$ and $\gamma = 0.99$.}
\image{width=0.5\columnwidth}{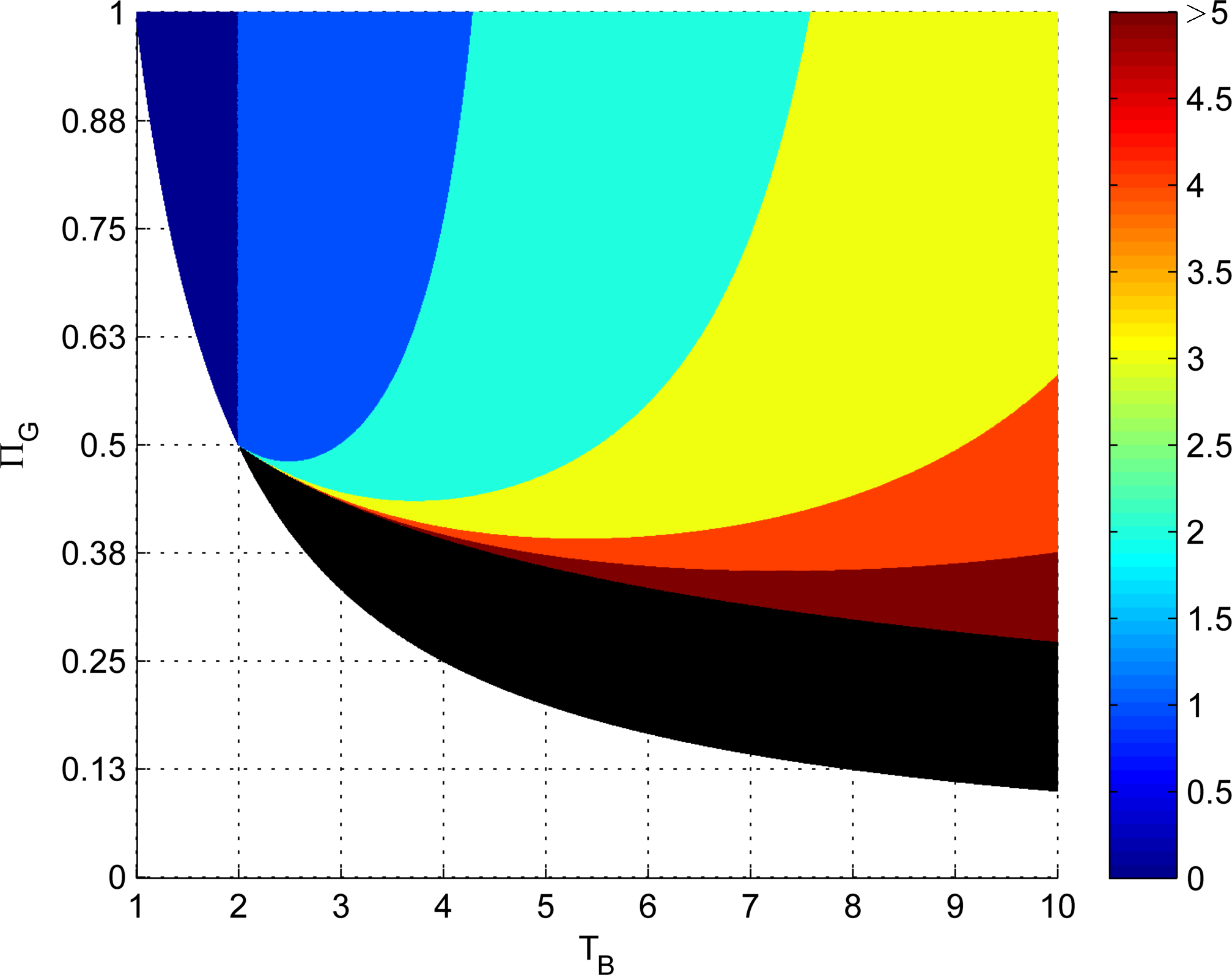}{Optimal sleeping time with $r_1 = 10$, $r_0 = 10$ and $\gamma = 0.99$.}
\image{width=0.5\columnwidth}{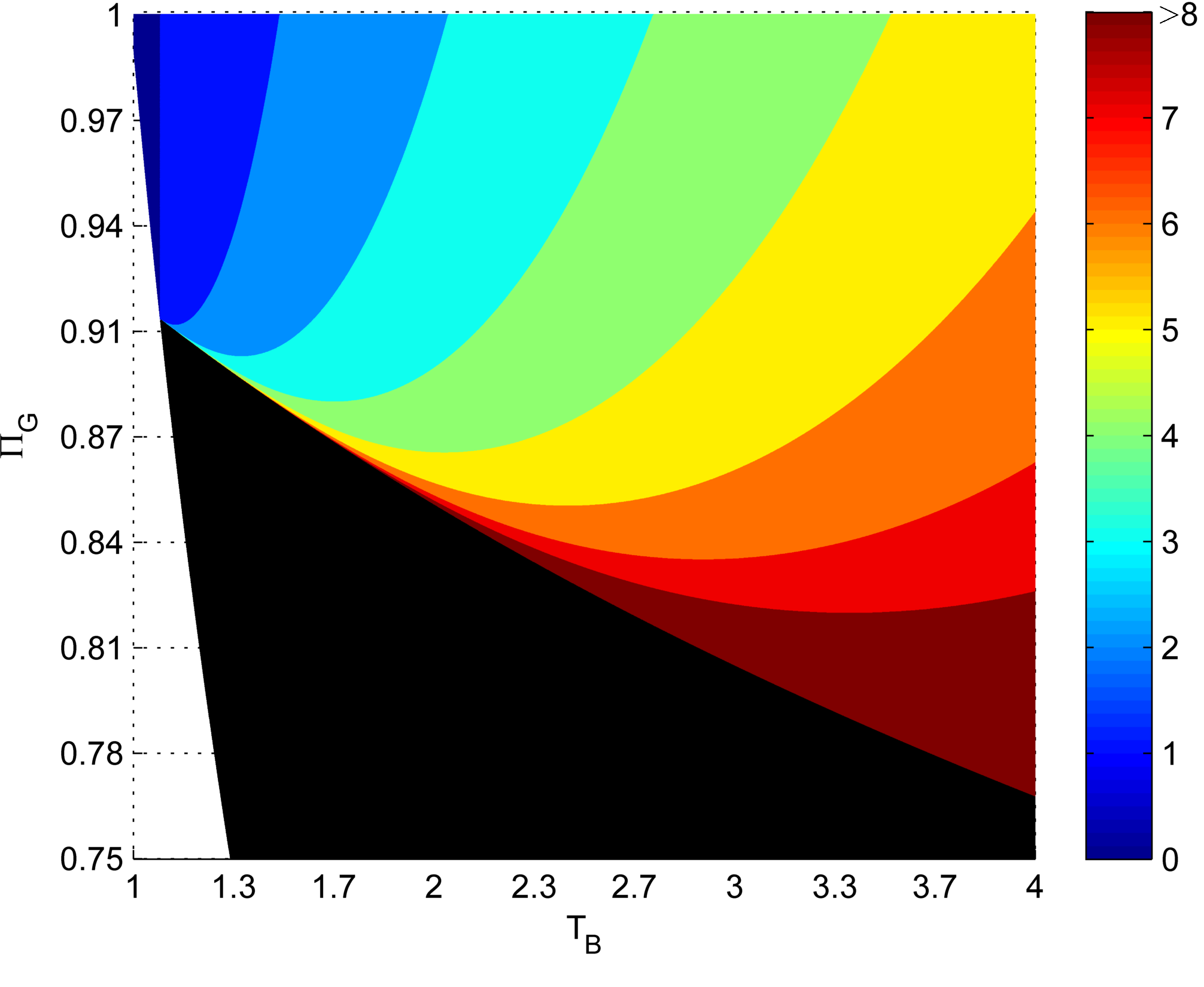}{Optimal sleeping time with $r_1 = 1$, $r_0 = 10$ and $\gamma = 0.99$.}

In the case of known transition probabilities of the Markov chain model, the optimal energy harvesting policy can be fully characterized by the sleeping time after an unsuccessful harvesting attempt (cf. Proposition~\ref{prop:waitTime}).
For different values of reward and cost, we show in Figure~\ref{fig:LookupTableHighRewards}--\ref{fig:LookupTableLowRewards} the optimal sleeping time, indexed by the average number of time slots the model stays in the bad harvesting state $T_B = 1 / q$ and the probability of being in the good state $\Pi_G = q / (p+q)$.
Note that the bottom-left region without any color corresponds to the case $1-p > q$.
The region with black color denotes the scenario in which it is not optimal to harvest any more after an unsuccessful harvesting.

From these figures, we first observe the natural monotonicity of longer sleeping time with respect to longer burst lengths and smaller success probabilities.
Moreover, the optimal sleeping time depends not only on the burst length and the success probability, but also depends on the ratio between the reward $r_1$ and the penalty $r_0$.
One might be mislead to believe that if the reward is much larger than the cost, then the optimal policy should harvest all the time.
However, Figure~\ref{fig:LookupTableHighRewards} shows that for a rather large parameter space, the optimal policy is to sleep for one or two time slots after an unsuccessful harvesting.
On the other hand, when the cost is larger (i.e.~larger $r_0$), it is better not to harvest at all in a larger parameter space.
Nevertheless, there still exists a non-trivial selection of the sleeping time to maximize the total harvested energy as shown in Figure~\ref{fig:LookupTableLowRewards}.  Figure~\ref{fig:maxEnergyLowReward} shows that the accumulated energy can be significant.
\image{width=0.5\columnwidth}{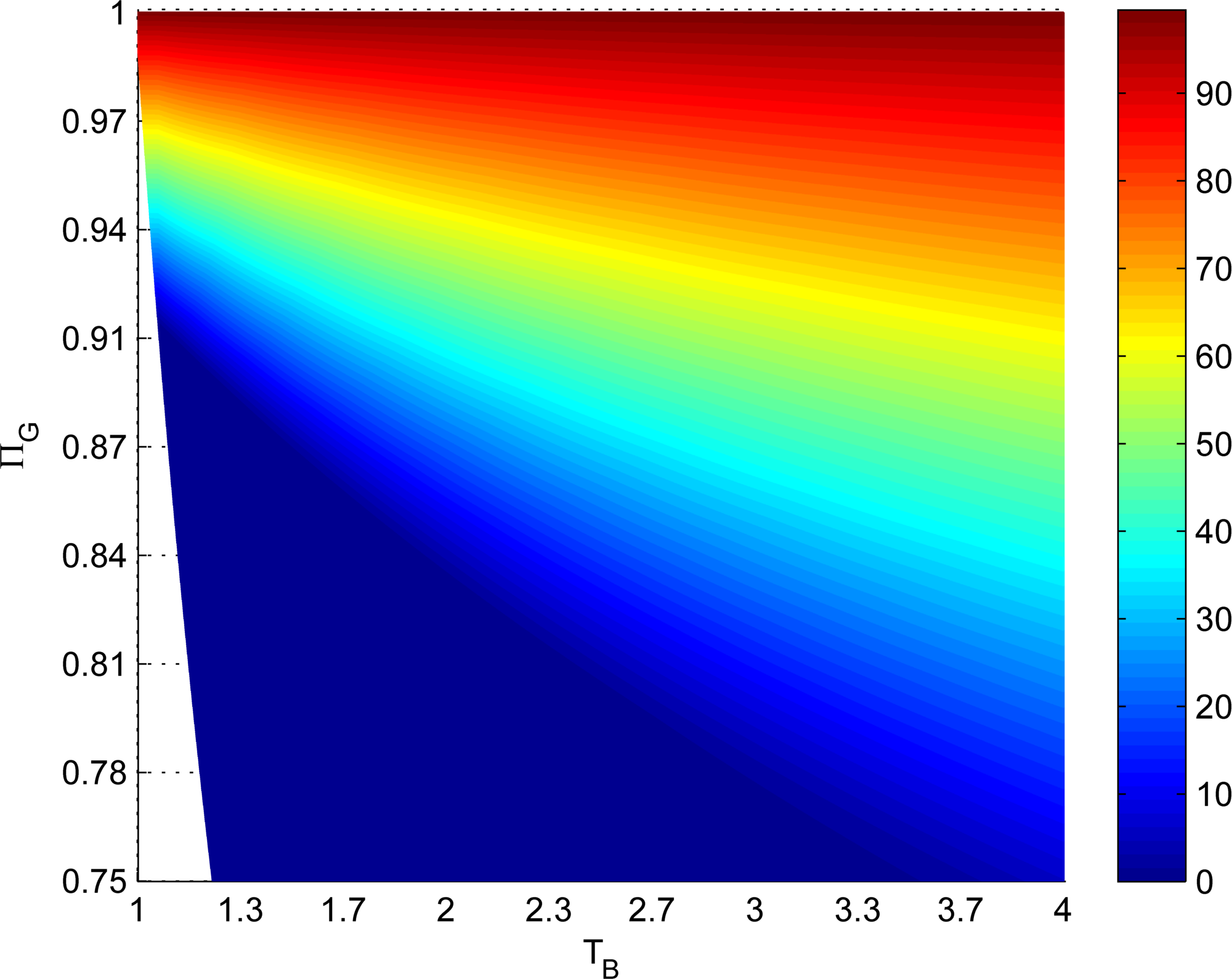}{Maximum harvested energy with $r_1 = 1$, $r_0 = 10$ and $\gamma = 0.99$.}

In these numerical examples, we let the reward $r_1$ and the penalty $r_0$ be close, and the ratio is between $0.1$ and $10$.
We believe such choices are practical.
For example, in AT86RF231~\cite{at86rf231802low} (a low power radio transceiver), it can be computed that sensing channel takes $3 \mu J$ energy since one clear channel assessment takes $140 \mu s$ and the energy cost for keeping the radio on is $22mW$.
Moreover, the energy harvesting rate of the current technology is around $200 \mu W$~\cite{Xiao2015Survey, popovic2013low}.
Suppose the coherence time of the energy source is $T$ milliseconds, which corresponds to the duration of the time-slot.
The ratio $r_1/r_0$ is roughly $(0.2 T - 3)/3$, and it ranges from $0.3$ to $10$ if $T \in [20,200]$ milliseconds.
Therefore, the ratio between the reward $r_1$ and the penalty $r_0$ is neither too large nor too small, and the POMDP and the threshold-based optimal policy is very useful in practice to derive the non-trivial optimal sleeping time.

Recall that the threshold-based optimal policy in Proposition~\ref{prop:waitTime} induces a discrete-time Markov chain with state $(S,\mathcal{E})$ which denotes the energy arrival state at the previous time-slot and the energy level at the current time-slot, respectively.
Note that, once the battery is completely depleted, we cannot turn on the radio to harvest anymore, which corresponds to the absorbing states $(S,0)$ for any $S$ in this Markov chain.
Suppose the maximum energy level is $\overline{\mathcal{E}}$, which introduces the other absorbing states $(S, \overline{\mathcal{E}})$ for any $S$.
Without loss of generality, we assume the energy level in the battery is a multiple of the harvested energy at each time-slot and the cost for an unsuccessful harvesting.
Hence, this Markov chain has a finite number of states, and we can derive some interesting parameters by standard analysis tools from the absorbing Markov chain theory~\cite{kemeny1960finite}.

Figure~\ref{fig:fullChargeProb.pdf} shows the full-charge probability under a hypothetical energy harvesting device with average success energy arrival probability equal $0.7$ and under different initial energy levels.
We assume that the maximum battery level is $100$ units, and one successful harvesting accumulates one unit of energy while one unsuccessful harvesting costs one unit of energy.
The plots can guide us in designing appropriate packet transmission policies.
For example, for the case of burst length equal $10$, we should restrain from transmitting the packet once the battery is around $20\%$ full if we want to keep the depleting probability smaller than $5 \cdot 10^{-4}$.
\image{width=0.6\columnwidth}{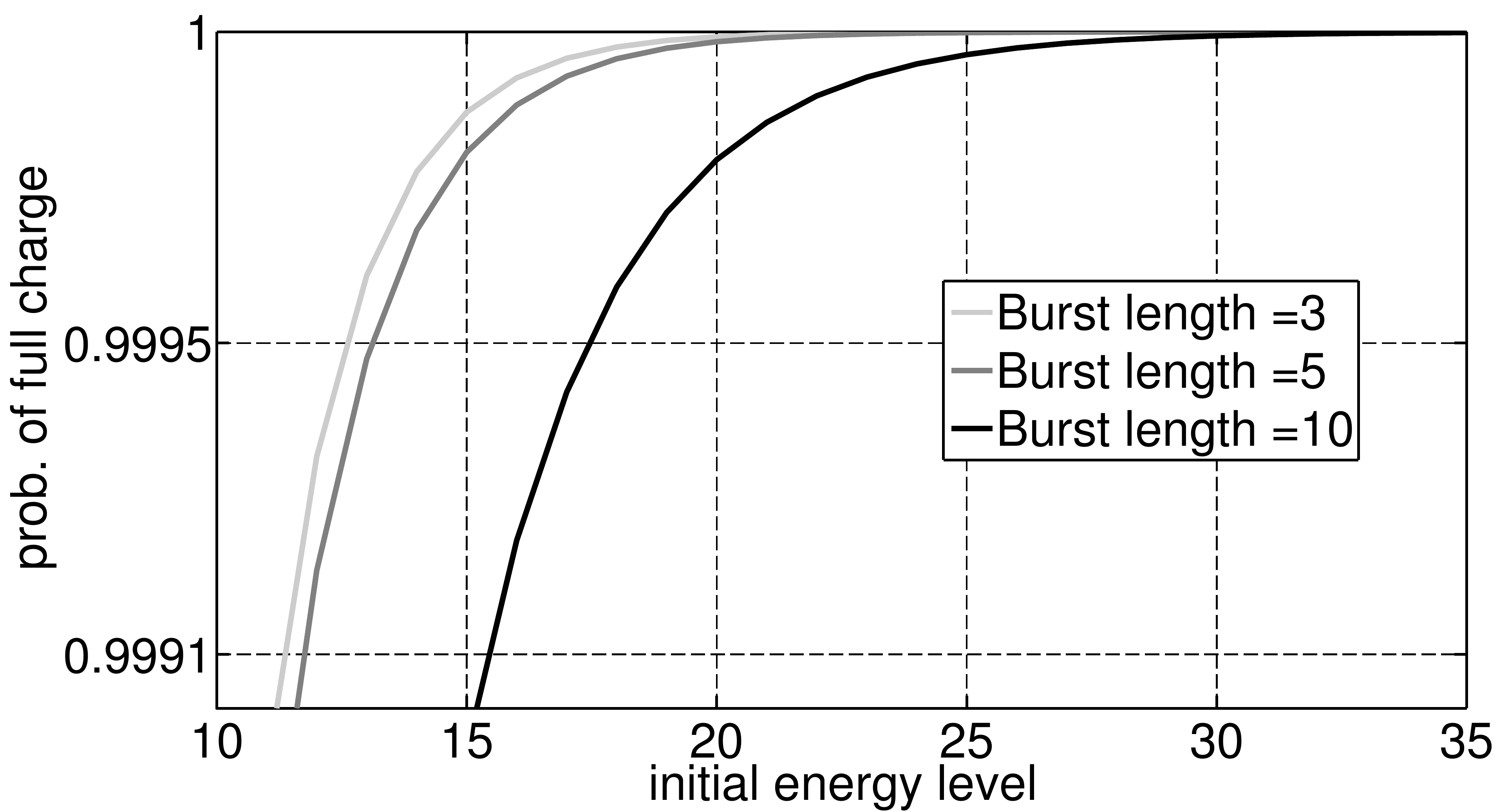}{The full-charge probability under different initial energy levels and average burst length.
}

Lastly, Figure~\ref{fig:WaitTime.pdf} shows the average number of time-slots to reach full-charge if the device manages to fully charge the battery, under different initial energy levels and average burst lengths.
The figure shows a decreasing and almost linear relation between the initial energy level and the average number of time-slots when the initial energy level becomes larger.
Similarly, the slope of these numbers can help us determine whether we can expect to be able to support a sensor application with a specified data transmission rate.
Suppose the cost for one packet transmission is $40$.
If the data rate is larger than one packet per $50$ time slots, the energy harvesting device would quickly deplete the battery, since it takes more than $50$ time slots to harvest $40$ units of energy.
On the other hand, if the data rate is smaller than one packet per $100$ time slots, then we are confident that it can support such applications.
\image{width=0.6\columnwidth}{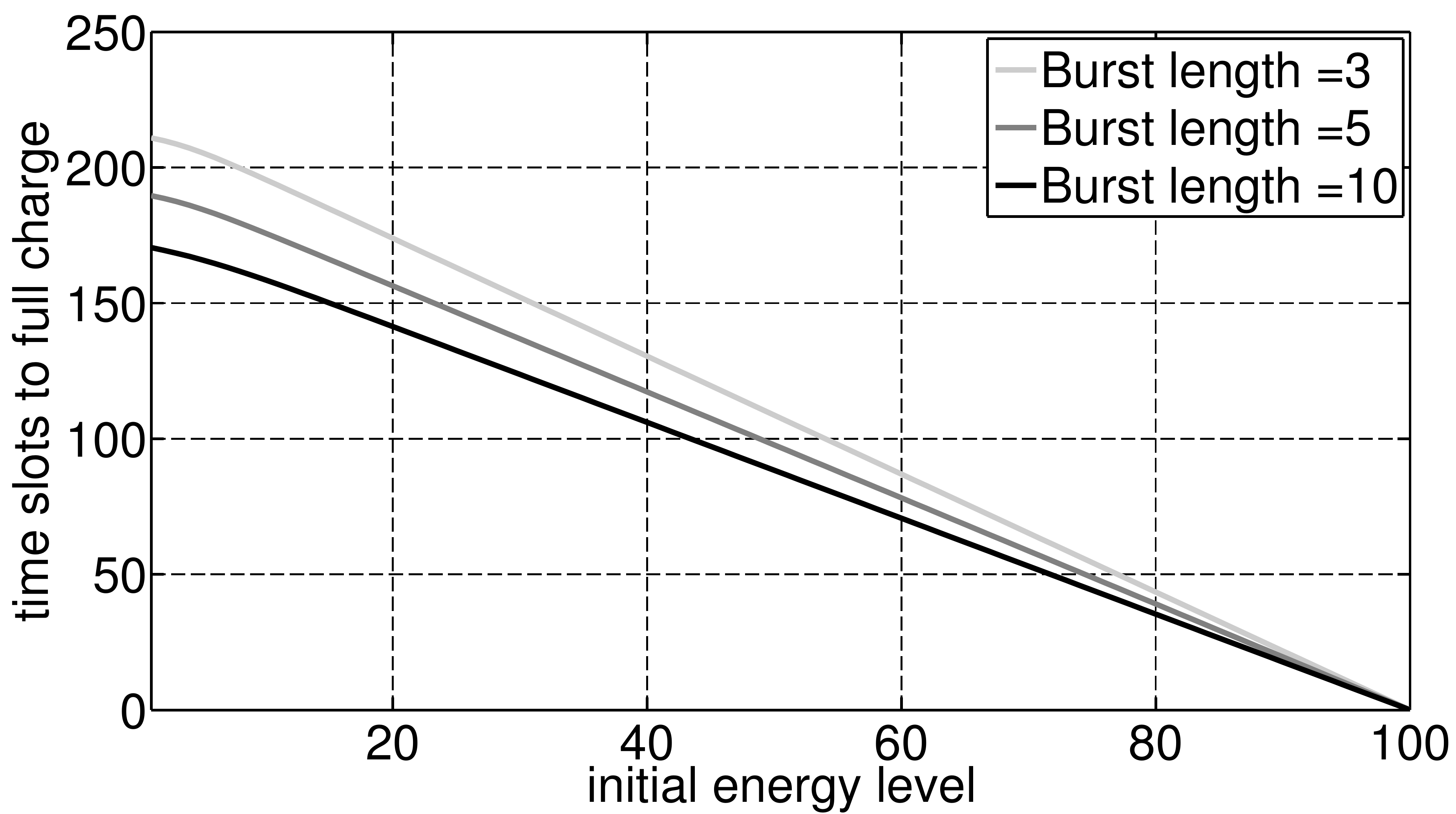}{The expected number of time-slots to reach full-charge under different initial energy levels and average burst length.
}

\subsection{Unknown transition probabilities}

In this section, we demonstrate the performance of the Bayesian learning algorithm.
Figure~\ref{fig:perform.pdf} shows that the performance of Algorithm~\ref{alg:Heuristic} outperforms other heuristic learning algorithms in terms of the total discounted reward.
The results are averaged over three hundred independent energy arrival sample paths generated from the unknown Markov chain, and for each sample path the rewards are averaged over one hundred independent runs.
In the heuristic posterior sampling method, the posterior count is only updated when we have an observation of the state transition (i.e.,~two consecutive harvesting actions that both reveal the state of the Markov chain).
In the heuristic random sampling method, we replace Line~\ref{algLine:PosteriorSample1}~and~Line~\ref{algLine:PosteriorSample2} in Algorithm~\ref{alg:Heuristic} with a uniformly selected set of parameters $p$ and~$q$.
Because of the heuristic choice of keeping only $K$ posterior counts, the Bayesian update is not exact and the parameter estimation is biased.
However, its total reward still outperforms others as a result of its smarter exploration decisions during the learning phase.
Note also that due to the discount factor $\gamma$ being strictly smaller than one, the reward and the penalty after five hundred time-slots are negligible compared to the already accumulated rewards.

\image{width=0.6\columnwidth}{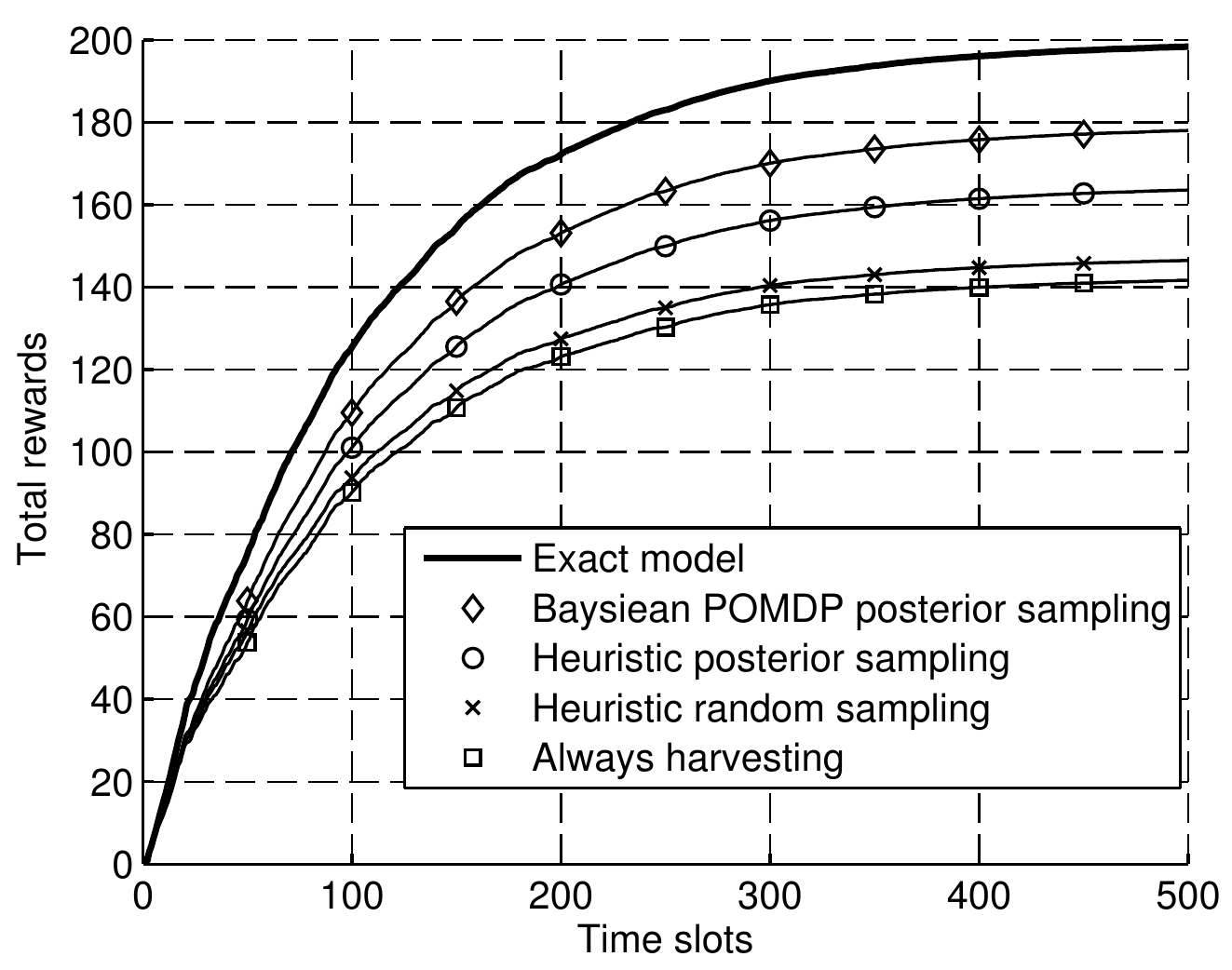}{Total rewards with different algorithms with $\Pi_G = 0.6$, $T_B = 2.5$, $r_0 = 10$, $r_1 = 10$, $\gamma = 0.99$, $K=20$.}

%
%
%
%
\section{Conclusions and Future Work}
\label{sec:conclusions}

\subsection{Conclusions}
In this paper, we studied the problem of when a wireless node with RF-EH capabilities should try and harvest ambient RF energy and when it should sleep instead.
We assumed that the overall energy level is constant during one time-slot, and may change in the next time-slot according to a two-state Gilbert-Elliott Markov chain model. Based on this model, we considered two cases: first, we have knowledge of the transition probabilities of the Markov chain. On these grounds, we formulated the problem as a Partially Observable Markov Decision Process (POMDP) and determined a threshold-based optimal policy. Second, we assumed that we do not have any knowledge about these parameters and formulated the problem as a Bayesian adaptive POMDP. To simplify computations, we also proposed a heuristic posterior sampling algorithm. Numerical examples have shown the benefits of our approach.

\subsection{Future Work}

Since energy harvesting may result in different energy intakes, part of our future work is to extend the Markov chain model to account for as many states as the levels of the harvested energy and in addition to include another Markov chain that models the state of the battery.

The problem of harvesting from multiple channels is of interest when considering multi-antenna devices. The formulation of this problem falls into the restless bandit problem framework and left for future work.

Finally, part of our ongoing research focuses on investigating what can be done when the parameters of the Markov chain model change over time.

\bibliographystyle{IEEEtran}
\bibliography{references}

%
%
%
%
\end{document}